\documentclass[twocolumn,showpacs,pre,floatfix,superscriptaddress]{revtex4-1}
\usepackage{subfigure}
\usepackage[utf8]{inputenc}
\usepackage{comment}
\usepackage{color} 
\usepackage{tabularx} 
\usepackage{epsfig}
\usepackage{amsmath} 
\usepackage{amssymb} 
\usepackage{graphicx}
\usepackage{wasysym}
\usepackage{oldgerm}
\usepackage{psfrag}
\usepackage{placeins}
\usepackage{setspace}

\usepackage{url}
\newcommand{\WP}{Wikipedia page-views}
\newcommand{\pw}{page-views}

\begin{document}

\title{Untangling Performance from Success}

\author {Burcu Yucesoy }
\affiliation{Center for Complex Network Research and Department of Physics, Northeastern University, Boston,  USA}
\author {Albert-L\'aszl\'o Barab\'asi }
\affiliation{Center for Complex Network Research and Department of Physics, Northeastern University, Boston,  USA}
\affiliation{Center for Cancer Systems Biology, Dana Farber Cancer Institute, Boston, USA}
\affiliation{Department of Medicine, Brigham and Women's Hospital, Harvard Medical School, Boston, USA}
\date{\today}

%\maketitle %\thispagestyle{empty} %% <-- you need this for the first page

%\date{\today}

\begin{abstract} 
Fame, popularity and celebrity status, frequently used tokens of success, are often loosely related to, or even divorced from professional performance. This dichotomy is partly rooted in the difficulty to distinguish performance, an individual measure that captures the actions of a performer, from success, a collective measure that captures a community's reactions to these actions. Yet, finding the relationship between the two measures is essential for all areas that aim to objectively reward excellence, from science to business. Here we quantify the relationship between performance and success by focusing on tennis, an individual sport where the two quantities can be independently measured. We show that a predictive model, relying only on a tennis player's performance in tournaments, can accurately predict an athlete's popularity, both during a player's active years and after retirement. Hence the model establishes a direct link between performance and momentary popularity. The agreement between the performance-driven and observed popularity suggests that in most areas of human achievement exceptional visibility may be rooted in detectable performance measures. 

\end{abstract}
%\pacs{?}

\maketitle

\section{Introduction}
\label{sec:Intro}
Performance, representing the totality of individual achievements in a certain domain of activity, like the publication record of a scientist or the winning record of an athlete, captures the actions of an individual \cite{lehmann2006measures, paper:radicchi2008universality, paper:Grubb1998, paper:Radicci2011, paper:Radicci2012}. In contrast success, fame, celebrity, or visibility are collective measures, representing a community's reaction to and acceptance of an individual's performance \cite{paper:uzzi2008social, paper:ke2015defining, paper:yu2015pantheon}. The link between these two measures, while often taken for granted, is actually far from being understood and often controversial and lopsided. Indeed, even the most profound scientific discovery goes unnoticed if its importance is not acknowledged through discussions, talks and citations by the scientific community. The void between success and performance is well illustrated by the concepts of `famesque', `celebutante' or `faminess', used to label an individual without tangible performance, but ``known for his well-knowingness" \cite{book:Boorstin2012}. These often prompt us to see fame and success as only loosely related to~\cite{paper:van2013only, gabler2001toward, andrews2002sport} and often divorced~\cite{article:Famesque, whannel2013media, book:turner2004,  Book:evans2005, ivaldi2008adolescents, andrews2002sport} from performance. This dichotomy is illustrated by documented examples of scientists whose popular media visibility significantly exceeds their scientific credentials ~\cite{paper:hall2014kardashian}, or by countless celebrities, from the Kardashian sisters to athletes with no or only underwhelming accomplishments \cite{article:kardashian, web:Looks, web:Nothing}, as well as by high performers like David Beckham or Tiger Woods who are frequently featured in the media for reasons unrelated to their professional achievements~\cite{paper:Beckham, essay:Woods}.  
The source of this dichotomy is that in most areas of human achievement it is difficult to distinguish performance from success \cite{book:murray2003human}. Indeed, while we can use citations, prizes and other measures to quantify the impact of a scientific discovery, we lack objective performance measures to capture the degree of innovation or talent characterizing a particular paper or a scientist. 

Our goal here is to explore in a quantitative manner the relationship between performance, an individual measure, and success, a collective measure capturing the societal acknowledgement of a given level of performance. We do so through sports, an area where performance is accurately recorded in terms of number of wins, place in rankings or career records~\cite{paper:Sire2009, merritt2014scoring, paper:Clauset1, paper:lago2010, paper:Petersen2011, paper:Motegi2012, paper:VazdeMelo2012}. Sports is characterized by an equally obsessive focus on popularity and fame, which strongly affects an athlete's market value \cite{Herm2014484}.

Yet, in sports too, performance and success often follow different patterns, illustrated by the fact that only a small fraction of the earnings of a professional athlete is tied directly to his/her performance on the field, the vast majority coming from endorsements, determined more by the athlete's perceived success. 
For example, only \$4.2 million of Roger Federer's \$56.2 million reported 2014 income was from tournament prizes \citep{web:Forbes}, the rest came from endorsements tied to the public recognition of the athlete. 
Yet, Novak Djokovic, who was better ranked than Federer during 2014, received over \$12.1 million prize money but only \$21 million via endorsements, about a third of Federer's purse.
The fact that professional performance does not uniquely determine reward is further illustrated by Anna Kournikova, who in 2003 was the second best paid female tennis player, despite never reaching higher than the 8th place in rankings. This and many other well documented cases of "faminess" raise an important question: What performance factors affect popularity and how do they do so? In other words, can performance explain popularity and fame and if so, to what extent? These are fundamental issues in most areas that aim to fairly reward excellence, from science to education and business.

\begin{figure*}%[!h]
\center
\includegraphics[width=2.\columnwidth]{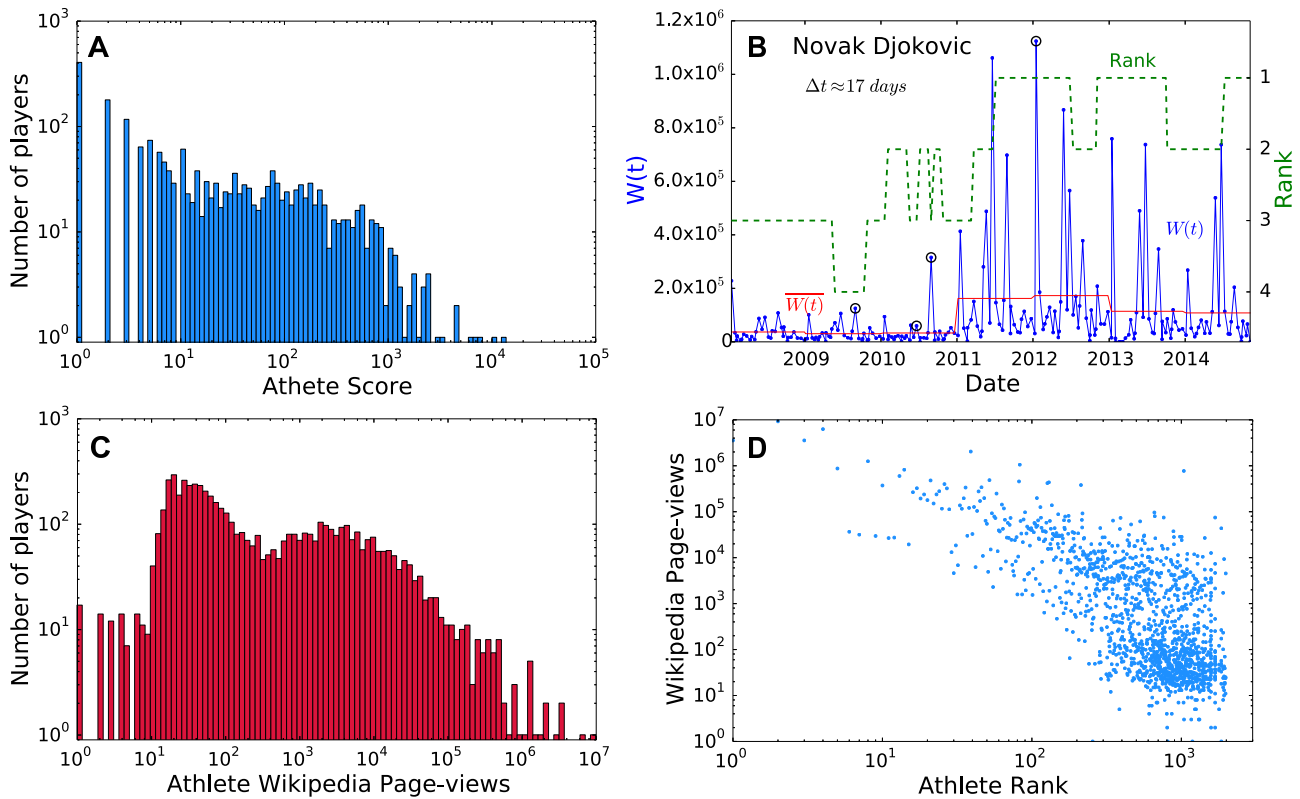} 
\caption{ \textbf{Ranking and visibility.}
(\textbf{A}) The score distribution for the players on the ATP rankings list on December 31, 2012.
(\textbf{B}) Page-views $W(t)$ (blue) and the rank (green) for Novak Djokovic, where the time $t$ corresponds to either the beginning of a tournament or every 17 days, a period slightly longer than the average duration of a Grand Slam. The red line indicates $\overline{W(t)}$, the yearly average \pw. The four marked data points correspond to the 2009 semi finals ($n(t)=6$) of US Open ($V(t)=2,000$) when Djokovic lost against the top ranked Federer ($\Delta r(t) H(\Delta r)/r(t)=1/4$), gaining him the highest visibility peak up to that point, the semi finals ($n(t)=4$) at Shanghai Masters ($V(t)=1,000$) in 2010 for which $H(\Delta r)=0$; winning ($n(t)=7$) the US Open ($V(t)=2,000$) against Nadal ($\Delta r(t) H(\Delta r)/r(t)=2/3$), which lead to an explosion in Djokovic's popularity. His tallest peak was in 2012 when he won the Australian Open ($V(t)=2,000$) while he was at the top of the rankings.
(\textbf{C}) The distribution of the number of \WP\ for athletes active and retired during the year 2012.
(\textbf{D}) Number of \WP\ for all active players during the year 2012 shown in function of their ATP ranking on December 31, 2012. (Also see Fig. S1 for page-view counts of active players and how those relate to their score points).
\label{fig:RankVis}}
\end{figure*}

\section{Performance and Success In Tennis}
\label{sec:Main}
Professional tennis players accumulate score points based on their winning record during the previous year (Fig.~\ref{fig:RankVis}A). By ordering the players based on their score we obtain the official Association of Tennis Professionals (ATP) Rankings, $r(t)$, an accurate measure of a tennis player's performance relative to other players, having $r=1$ for the top player while $r$ is high for low-performing athletes. 

We measure each player's time-resolved visibility through the number of hourly visits to its Wikipedia page \cite{paper:Wikipopular, paper:Wiki, paper:keegan2013hot}, a sensitive time-dependent measure of the collective interest in a player (Fig.~\ref{fig:RankVis}B). We define a player's popularity or fame through his cumulative visibility, representing the average number of visits his Wikipedia page acquires during a year (red line in Fig.~\ref{fig:RankVis}B).
Figure~\ref{fig:RankVis}C indicates that top players can gather a total of $10^7$ visits in a year, while those at the bottom of the rankings collect as few as $10$, documenting popularity differences between players that span orders of magnitude. %Regardless of their status (active or not), most players received a few \pw\ per day. A few players, however, accumulate orders of magnitude more \pw\ ($>10^4$). 
As Figure~\ref{fig:RankVis}D indicates, a player's rank (performance) and Wikipedia visits (popularity) are correlated: The lower the ranking, the higher are the \WP. Yet, we also observe a significant scattering: Athletes ranked around $r=$1,000 can gather anywhere from 10 to $10^5$ visits per year, wide differences that support the impression that popularity is often divorced from performance. 

\begin{figure*}
\center
\includegraphics[width=2.\columnwidth]{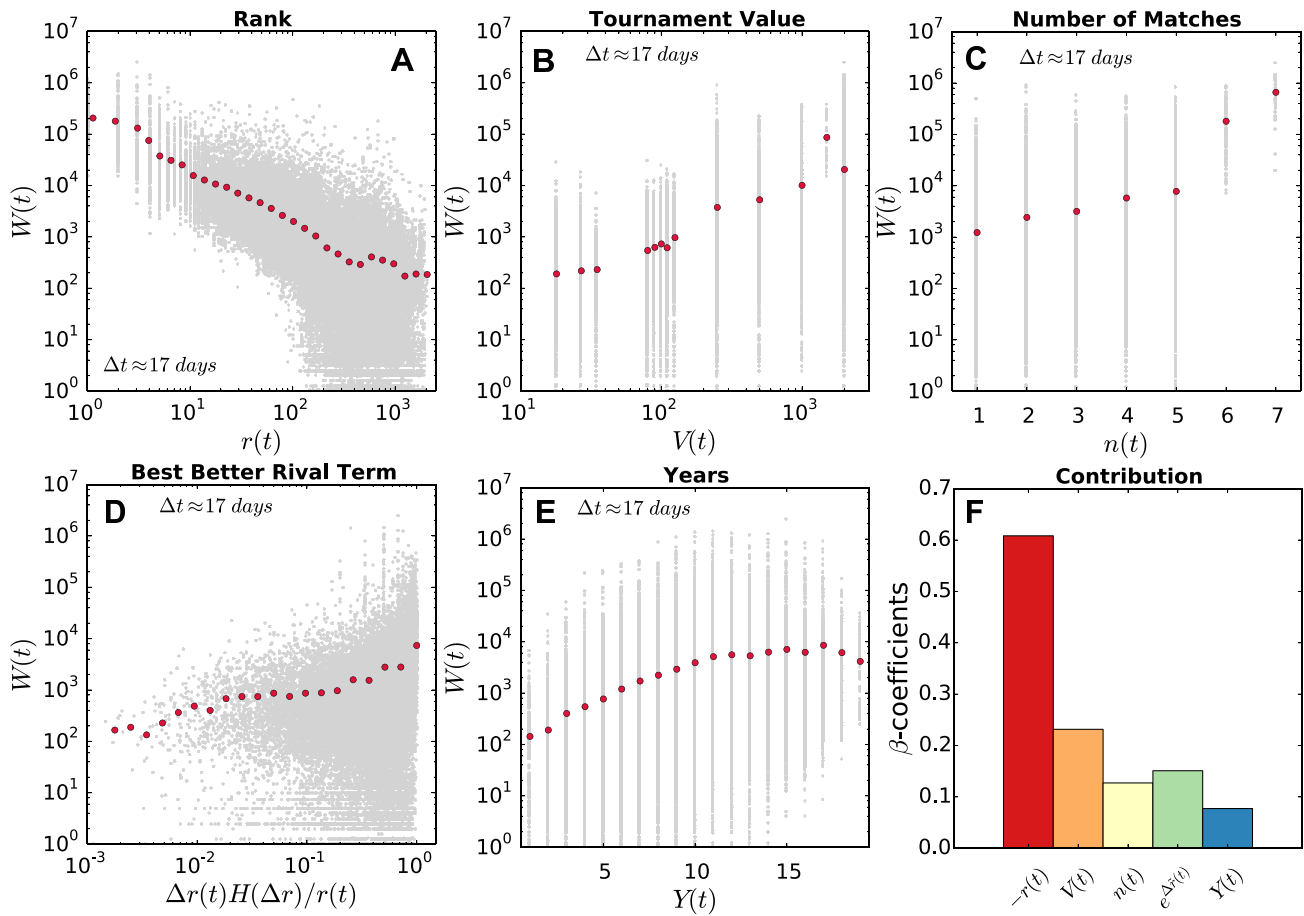} 
\caption{ \textbf{The impact of performance on visibility.} The \WP\ $W(t)$ for all players vs. the various measures of individual performance: (\textbf{A}) the rank of a player $r(t)$; (\textbf{B}) the tournament value $V(t)$; (\textbf{C}) the number of matches the player participates in during a tournament $n(t)$; (\textbf{D}) the rivals term $\Delta r(t) H(\Delta r) /r(t)$; (\textbf{E}) the number of years active $Y(t)$. The Wikipedia visits, each corresponding to $\Delta t\approx 17$ day bins, are shown in grey while the red dots correspond to binned averages. 
In (C), up to $n(t)=5$ (where most tournaments end) $W(t)$ increases with $n(t)$ (Fig.~\ref{fig:IoP}C). The jump for $n(t)=6$ and $7$ corresponds to the semi-final and final matches of the most watched tennis tournaments, offering disproportionately more visibility. In (E), after 15 years the visibility drops slightly, indicating that players with very long professional career no longer benefit from career longevity. 
(\textbf{F}) The contribution of each performance variable to visibility $W(t)$ based on (\ref{eq:Model}). Here $\Delta \tilde{r}(t)\equiv \Delta r(t) H(\Delta r) /r(t)$ and the $\beta$-coefficients result from a multivariate regression analysis of the form $log\ y=\sum_i a_i\ log\ x_i+c$. The dependent variable $y$ is the Wikipedia page-view $W(t)$; $x_i$ represents one of the five independent variables ($r(t),\ V(t),\ n(t),\ e^{\Delta r(t) H(\Delta r) /r(t)},\ Y(t)$), and the $a_i$ are the corresponding coefficients. All variables are individually significant ($p<0.001$).
\label{fig:IoP}}
\end{figure*}

To address the degree to which performance determines popularity, we reconstructed the number of Wikipedia visits $W(t)$ for all ranked professional male tennis players between 2008 and 2015 (Fig.~\ref{fig:RankVis}B), together with their time-dependent history of achievements on the field (supplementary material S3). The two datasets allow us to identify several performance-related factors that influence a player's visibility, and eventually his popularity:

(i) \textit{Rank $r(t)$}: Fig.~\ref{fig:IoP}A shows the measured \WP\ $W(t)$ vs. a player's momentary rank $r(t)$, indicating that the number of visits rapidly drop with the increasing rank of a player. It also shows that the variations in $W(t)$ is higher for players with lower performance (i.e. higher rank) (Fig.~\ref{fig:RankVis}B). 

(ii) \textit{Tournament value $V(t)$}: The more points $V(t)$ a tournament offers to the winner, a measure of the tournament prestige, the more visibility it confers to its players, win or lose (Fig.~\ref{fig:IoP}B). For example, the early peaks in Djokovic's visibility correspond to his participation in the US and Australian Opens, two high value tournaments (Fig.~\ref{fig:RankVis}B).
 
(iii) \textit{Number of matches $n(t)$}: The more matches an athlete plays within a tournament, reflecting his advance within the competition, the more exposure he receives (Fig.~\ref{fig:IoP}C). As the losing player is eliminated, $n(t)$ also determines the amount of points the player collects within a tournament. 
%THIS INTO CAPTION: Up to $n(t)=5$ (where most tournaments end) we see a steady increase of $W(t)$ with $n(t)$ . The jump for $n(t)=6$ and $7$ exist only for highly valued tournaments, corresponding to the semi-final and final matches in the most watched tennis tournaments, offering disproportionately more visibility. 

(iv) \textit{Rank of the best opponent}: A match against a better ranked rival generates additional interest in a player. To capture this effect we measure the relative rank difference for each match $\Delta r(t) H(\Delta r) /r(t)$, where $\Delta r(t)=r(t)-r_{BR}(t)$ is the difference between the rank of the considered player and his best rival in the tournament; the Heaviside step function $H(\Delta r)$ is one when the opponent has a better rank and zero otherwise. The increase of the average \pw\ with $\Delta r(t) H(\Delta r) /r(t)$ quantifies the boost in visibility from playing against a better athlete (Fig.~\ref{fig:IoP}D), like the visibility peak of Djokovic when he played against the then \# 1 Federer in the 2009 US Open (Fig.~\ref{fig:RankVis}B) (see S9 for the detailed opponent statistics).

(v) \textit{Career length $Y(t)$}: The longer the player has been an active professional player, the more Wikipedia visitations he collects (Fig.~\ref{fig:IoP}E). %After 15 years the effect slightly reverses, indicating that players with a very long professional career no longer benefit from career longevity. 

Figures~\ref{fig:IoP}A -- E document clear correlations between the performance measures (i)-(v) and the momentary visibility of a player. Yet, a linear fit $y(t)=Ax(t)+C$ of each individual performance measure to the observed \WP\ results in $R^2 < 0.1$, except for $1/r(t)$, for which $R^2=0.29$. Therefore, no individual performance measure can fully explain visibility, indicating that performance drives popularity through a combination of performance measures (see supplementary materials S4 for variable interdependencies). 
We therefore explored the predictive power of the sum of these variables with multipliers obtained via an ordinary least squares (OLS) fitting \cite{book:cohen2013applied} process, resulting in $R^2=0.31$, only slightly better than $1/r(t)$ alone. We find, however, that a multiplicative process offers a much better predictive power, an OLS fitting leading to the model \begin{equation}
W_M(t)=A \frac{1}{r(t)} V(t) n(t) e^{\frac{\Delta r(t) H(\Delta r)}{r(t)}} Y(t) + C, 
\label{eq:Model2}
\end{equation}
which yields $R^2=0.57$. This indicates that the influence of the performance measures (i)-(v) do not add up, but amplify each other in a multiplicative manner, allowing for the emergence of the extreme fluctuations in visibility, as observed in Fig.~\ref{fig:RankVis}B and Figs.~\ref{fig:IoP}A -- E.
By taking the logarithm of (\ref{eq:Model2}) and calculating the standardized $\beta$-coefficients for each term, %using the ratio of the standard deviations of the corresponding variable and the dependent variable. 
we can evaluate how strongly each performance variable influences $W(t)$ in units of standard deviation. 
Figure~\ref{fig:IoP}F shows the obtained standardized $\beta$-coefficients, indicating that rank is the strongest driving force of visibility, followed by the value (prestige) of the tournaments, the rank of the opponents and the number of matches the player participated in a tournament. While career length contributes to a lesser degree, all terms are significant ($p<0.001$).

These results lead to a popularity model (PROMO), which predicts the time dependent visibility of a player $W_M(t)$ using the athlete's performance as an input, 
\begin{equation}
W_M(t)=A \frac{Y(t)}{r(t)} V(t)n(t)\ e^{\Delta r(t) H(\Delta r)/r(t)} + C \frac{Y(t)}{r(t)}.
\label{eq:Model}
\end{equation}
The last term accounts for periods when the player is not playing (between tournaments, periods of injury, etc.). %On those cases both $V(t)$ and $n(t)$ are zero but the player still collects \pw, whose magnitude depends on the player's current rank and career length. 
An OLS fitting process of this two-parameter PROMO results in $R^2=0.60$, offering the best predictive accuracy of the models we tested. Hence (\ref{eq:Model}) represents our main result, linking an athlete's visibility ($W(t)$) to his performance captured by $r(t),\ V(t),\ n(t),\ \Delta r(t) H(\Delta r)/r(t)$ and $Y(t)$.

\begin{figure*}%[!h]
\includegraphics[width=2.\columnwidth]{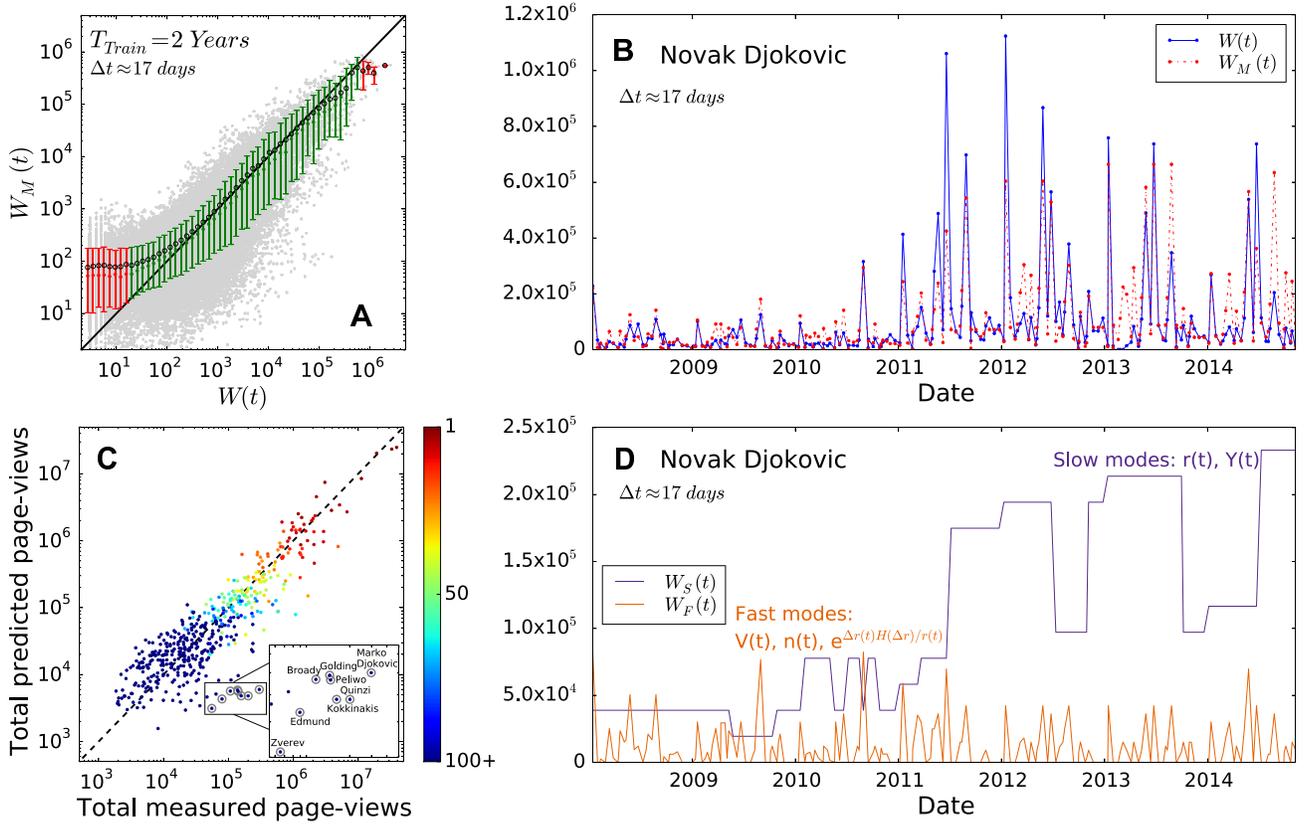} 
\caption{ \textbf{Predicting visibility and popularity.} 
(\textbf{A}) Scatter plot of predicted \WP\ $W_M(t)$ and collected \pw\ $W(t)$, using the parameters $A=3.747$ and $C=7929$ determined from the page-view history for the first 2 years (training data). Error bars indicate prediction percentiles (10\% and 90\%) in each bin and are green if $y=x$ lies between the two percentiles in that bin and red otherwise. The black circles are the average predicted \pw\ in that bin.
(\textbf{B}) Comparison between Novak Djokovic's observed \WP\ $W(t)$ (blue) and his performance-based predicted \pw\ $W_M(t)$. The model accurately captures the considerable lift in his career in 2011, when his \pw\ increased about an order of magnitude.
(\textbf{C}) The total predicted \WP\ for each player, capturing a player's predicted popularity, compared to the actual total \WP\ in the whole period considered in our analysis. The symbol colors represent the best rank the player reached during the considered period. The highlighted points correspond to the most significant outliers, whose modified $z$-value for the logarithmic distance between the prediction and data exceeds $3.5$. The corresponding athletes are identified in the insert.
(\textbf{D}) Separating the slow modes $W_S(t)$, driven by career length $Y(t)$ and rank $r(t)$ (purple), and fast modes $W_F(t)$, driven by tournament value $V(t)$, number of matches $n(t)$ and best better rivals term $e^{\Delta r(t) H(\Delta r) /r(t)}$ (orange).
\label{fig:Model}}
\end{figure*}

We designed several tests to validate PROMO's predictive power:

(i) In a prospective study we fit the r.h.s of (\ref{eq:Model}) to $W(t)$ for 2008 and 2009, obtaining the coefficients $A=3.747$ and $C=7929$, the same for all athletes (see supplementary materials S3). 
We then use these parameters to predict the momentary visitations $W_M(t)$ for the subsequent 5 years (Fig.~\ref{fig:Model}A). %The error bars are colored green if the y=x line lies between the 10\% and 90\% percentiles and red otherwise, indicating 
We find that the model accurately predicts the bulk of the real data. The deviation for very low \WP\ has a simple technical reason: beginning athletes often lack a dedicated Wikipedia page, hence their \pw\ are counted only indirectly and incompletely on Wikipedia (see supplementary materials S3 and S6). Once a player gets a Wikipedia page, his visibility approaches the model-predicted values. 

(ii) Figure~\ref{fig:Model}B compares the time dependent visibility $W_M(t)$ to the real visitation $W(t)$ for Novak Djokovic, indicating that the model (\ref{eq:Model}) accurately captures not only the explosion of his overall popularity in 2011 following his exceptional performance on the field, but also the peaks in the visitation patterns.

(iii) The model (\ref{eq:Model}) allows us to separate the role of the different parameters: Rank and career length act as \textit{slow modes}, driving the \textit{popularity} of an athlete, capturing his overall fame or celebrity (Fig.~\ref{eq:Model}D. In contrast the tournament value $V(t)$, the number of matches played within a tournament $n(t)$, and the rank of the best opponent represent \textit{fast modes} that drive the \textit{momentary visibility}, like the timing and the height of the individual visibility peaks (Fig.~\ref{fig:Model}D). As the slow and fast modes get multiplied, they together can account for the major visibility peaks on a slowly varying background, which in turn determines an athlete's overall fame.

(iv) To assess our ability to predict a player's popularity or fame from his performance, we use the total \pw\ a player received across several years. %like the number of views a player's Wikipedia page receives in a year. Figure~\ref{fig:YearlyA} and D show the comparison between the predicted and the actual yearly \pw\ for all players for all years. Each dot is a year in a player's career, the black circles/dots are averages over the predicted \pw\ in logarithmic bins, finding that the prediction for total yearly \pw\ is accurate for most players and years. In D we color-coded the data points to show the year they correspond to, indicating that data points farthest from the $x=y$ line generally stem from earlier years. Consequently as the total reach and penetration of Wikipedia grows, \WP\ becomes a better proxy for measuring popularity and the accuracy of our model increases. The relative popularity of players is captured by their total \pw. 
Figure~\ref{fig:Model}C shows a comparison between the predicted and observed popularity of each active player between 2008 and 2015, indicating that the observed popularity $W(t)$ closely follows the performance-driven predicted popularity $W_M(t)$. 
A color-coding by the player's peak rank %during the considered period 
reveals that for players that reached top rankings, the accuracy of the prediction is remarkable; scattering is only seen for lower ranked players. 

To understand if popularity in tennis can be induced by factors unrelated to performance, we inspected the outliers, athletes whose observed popularity is significantly higher than their performance-based popularity. We find that the outliers highlighted in Fig.~\ref{fig:Model}C (also listed in Table S3) are young players at the bottom of the rankings, who participated in only a few tournaments. An inspection of their career reveals that their added popularity is also performance driven, routed in outstanding results in junior or doubles tournaments, performance factors not considered in (\ref{eq:Model}).
For example Quinzi, Peliwo, Zverev and Golding reached number one or two in junior rankings, and Broady and Edmund had considerable successes in doubles. Only Marco Djokovic's visibility could not be explained by his performance. His higher than expected fame is most likely related to the attention (and potential confusion) he earns as the brother of Novak Djokovic, one of the best active tennis players. 

\begin{figure*}%[!h]
\includegraphics[width=2.\columnwidth]{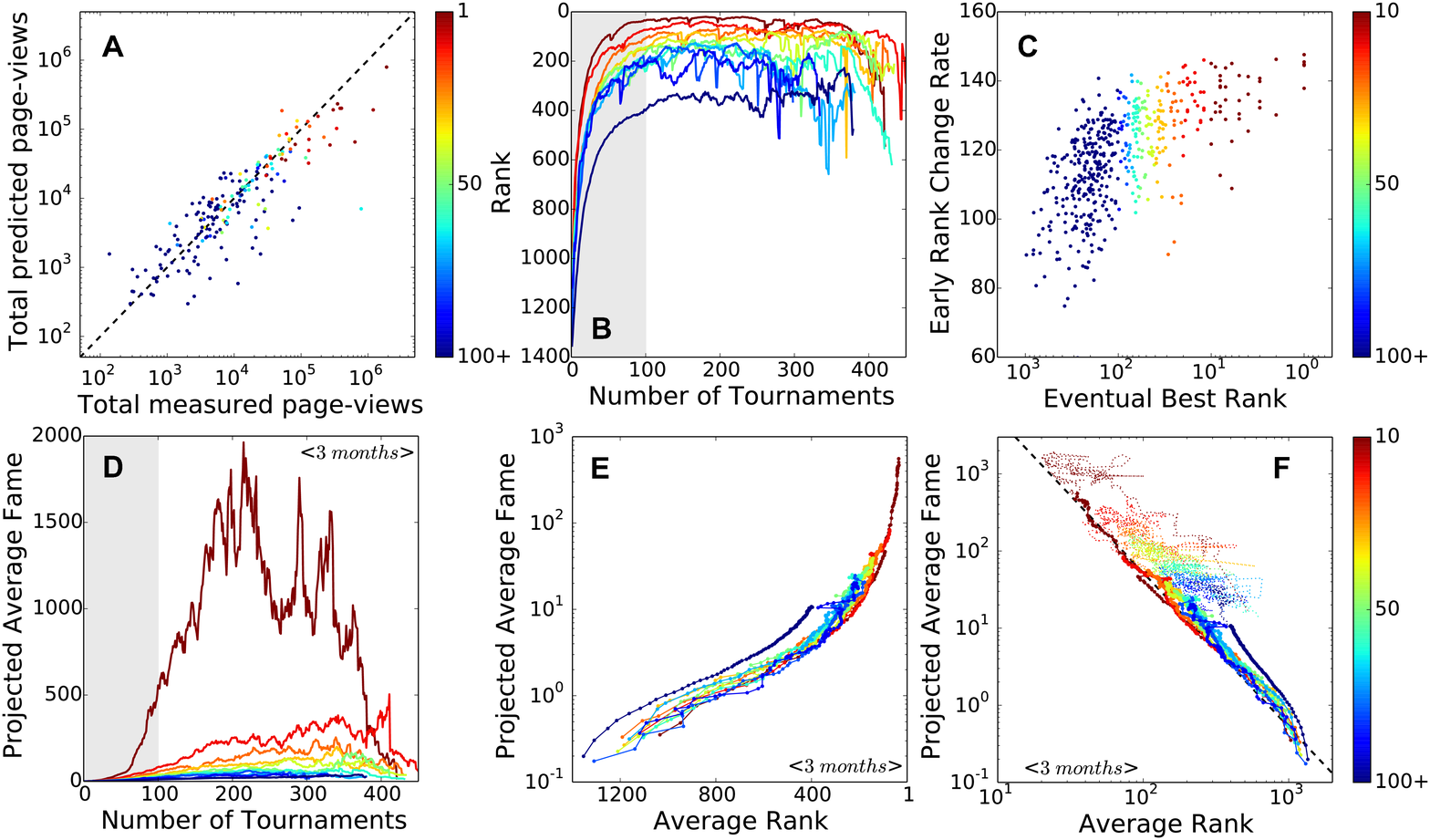}
\caption{ \textbf{The emergence of fame.} (\textbf{A}) The total predicted \WP\ for each retired player compared to the actual total \WP\ they collected after they stopped participating in tournaments, color-coded based on the best rank the player reached during his active career.
(\textbf{B}) The evolution of player rankings vs. the number of tournaments they participated in, shown as averages over categories based on the best ranking an athlete achieved during his career. Dark red captures top 10 players and dark blue those whose best ranking is more than 100. 
(\textbf{C}) The change in the rankings for the first 20 tournaments for all players plotted with respect to their eventual best rank.
(\textbf{D}) The projected daily average \pw\ of players based on a moving sum of three months preceding each tournament, averaged over categories based on best rank.
(\textbf{E}) The projected average \pw\ of players vs. their average rank at the time of the tournaments for the first 100 tournaments of their career, again averaged over categories based on best rank.
(\textbf{F}) The projected average \pw\ of players vs. their average rank at the time of the tournaments, for the duration of their careers. The first 100 tournaments are highlighted and the black dashed line represents a $<r>^{-2}$ fit. 
\label{fig:Dynamics}}
\end{figure*}

(v) Finally, we find that the Wikipedia pages of retired players continue to attract visitors (Fig.~\ref{fig:RankVis}C), prompting us to ask if a player's enduring popularity can be explained by his past performance. Given that for retired players $r(t)$, $V(t)$, $n(t)$ and $\Delta r(t) H(\Delta r)/r(t)$ are not recorded, their visibility can be determined only by the second term in (\ref{eq:Model}). By using the median for $r(t)$, reflecting a player's overall performance during his active career, and $Y_T$ for the number of active years $Y(t)$, we predict an athlete's popularity during retirement as
\begin{equation}
W^{Inact}_M(t)=C \frac{Y_T}{r_{med}}.
\label{eq:Retired}
\end{equation} 
As Figure~\ref{fig:Dynamics}A shows, the predicted popularity of inactive players is in excellent agreement with the measured popularity, indicating that past performance is the main source of their enduring fame (see supplementary materials S7 for outliers). 

%Outliers are only seen at the lower ranks, representing less accomplished players whose popularity exceeds their performance.Yet, for most outliers, their added popularity is also determined by performance measures not documented in our dataset. 
Finally we can apply the tools and insights developed above to explore the emergence of fame. For this we group players into ten categories based on their peak ranking during their career. The change in the average ranking of each group over time (Fig.~\ref{fig:Dynamics}B -- C) indicates that players who reach the best ranks distinguish themselves in their first 20 tournaments, i.e. the climb in rank very fast early in their careers. This effect is particularly clear in Fig. \ref{fig:Dynamics}C, which shows the rate of change in rankings during the first 20 tournaments, indicating that players that eventually reach the top rise much faster early on than the rest. 
%Players that reach top 20 ranks follow closely behind the top 10 group, but for the top 30 and below the separation between categories disappear. 
Therefore, top ranks are not reached by a slow improvement in skill, but instead young players come in with a given skill set, some remarkable, others less so, and rapidly reach the vicinity of their skill-determined ranking level, where they fluctuate for most of their career (Fig.~\ref{fig:Dynamics}B).
 
Figure~\ref{fig:Dynamics}D shows the projected daily average Wikipedia visitation of players grouped based on their best rank (supplementary materials S8). 
It allows us to uncover a highly nonlinear relationship between rank and popularity (Fig.~\ref{fig:Dynamics}E): popularity raises fast immediately following a player's entry into the professional field, but this growth rate slows down between ranks 1,000 and 400. This is followed by an exploding popularity for the elite players, those that reach ranking 200 and below. In other words, elite players benefit from a disproportional popularity bonus, not accessible for other ranked players. Overall, we find a robust relationship between a player's average popularity and rank: At the beginning of a player's career popularity increases as $1/<r>^{\alpha}$, with $\alpha \approx 2.0\pm0.02$, indicating that in this early stages of an athlete's career rank is the most important determinant of popularity (Fig.~\ref{fig:Dynamics}F).
After the first 100 tournaments, however, the influence of rank on popularity is less pronounced and average popularity fluctuates in the vicinity of its top value.

%\FloatBarrier

\section{Conclusions}
While we would like to believe that fame, visibility and popularity are uniquely determined by performance, representing well-deserved recognition for some sustained or singular achievement, a significant body of media research indicates otherwise, suggesting that fame follows patterns on its own, divorced from talent or performance \cite{book:Boorstin2012, paper:van2013only, gabler2001toward, andrews2002sport,article:Famesque, whannel2013media, book:turner2004,  Book:evans2005, ivaldi2008adolescents, paper:hall2014kardashian,article:kardashian, web:Nothing, web:Looks, paper:Beckham, essay:Woods}. Here we aimed to quantify the relationship between performance and popularity in an area where these two quantities can be individually measured. We did so by constructing a model to predict a tennis player's visibility captured by his \WP, a proxy of the athlete's popularity and fame. 
Taken together, we find that in tennis a player's popularity and momentary visibility are uniquely determined by his performance on the court. The agreement is especially good for elite players. 
This indicates that for athletes exceptional performance offers exceptional visibility, a level that is hard to modulate by exogenous events.  
For less accomplished players we observe deviations from the performance-predicted popularity, suggesting that in this case visibility can be manipulated by exogenous events. It is comforting, however, that for most outliers the extra visibility can be explained by performance factors not considered in our model, like achievements in doubles or junior tournaments.
In short, the better the performance of an athlete, the more accurately it determines his popularity, and the lesser the role of exogenous factors. 
Finally, we find that the fame of retired players is also determined by their past performance, indicating that exceptional past performance can lead to a commensurate lasting legacy.

%The modeling framework also allows us to examine how players achieve fame as they progress in their careers. The uncovered fast early rise of top players in both performance and fame suggests that stars can be spotted early on. Most important, we identify a fame premium for the elite athletes, whose visibility skyrockets in a nonlinear fashion as they reach top rankings.
Our methodology can be readily generalized to other areas where performance and visibility can be independently measured. The generalization to sports like chess, table tennis, golf or car racing is straightforward as the ranking systems of these sports are similar to tennis. With careful adjustments it may also be appropriate for team sports, allowing us to systematically explore how the performance of a professional athlete is tied to his/her or the team's collective success. 

We would like to believe that in most areas of human achievement fame and visibility are determined by some underlying performance indicators. The scientific challenge, however, is to systematically separate performance from success, a fundamental goal from science to management.
The excellent agreement between the performance-based and the observed popularity documented here makes us wonder to what degree faminess is real, suggesting that outstanding fame and popularity may be rooted in performance measures that are perhaps not yet accessible to us.

%\FloatBarrier

\section{Acknowledgments}
We wish to thank Tam\'as H\'amori for his guidance toward a deeper understanding of the tennis world, Kim Albrect for helpful visualizations and other colleagues at the CCNR, especially those in the the success group, for valuable discussions and comments. This research was supported by Air Force Office of Scientific Research (AFOSR) under agreement FA9550-15-1-0077. 

%\section{Contributions}
%B.Y. and A.-L.B. designed the research. B.Y. analyzed the data and prepared the figures. B.Y. and A.-L.B. prepared the manuscript.

%\section{Competing interests}
%The authors declare no competing financial interests.

%\bibliography{/Users/burcu/Dropbox/WorkingPaper/references,comments}
\bibliography{ref2}

\end{document}